\documentclass[aps,prl,twocolumn,superscriptaddress]{revtex4}
\usepackage{graphicx,epsf}
\usepackage{amsmath,amssymb}

\usepackage{color}

\newcommand{\aio}{A$_2$IrO$_3$}

\newcommand{\w}{\omega}
\newcommand{\e}{\epsilon}
\newcommand{\HK}{\mathcal{H}_{\rm K}}
\newcommand{\HM}{\mathcal{H}_{\rm u}}

\newcommand{\HMh}{\mathcal{H}_{\rm \hat{u}}}

\newcommand{\ket}[1]{\left|#1\right\rangle}

\newcommand{\ELL}{\varepsilon}  
\newcommand{\EB}{E}             
\def\ie{\emph{i.e.},\ }

\def\etal{\emph{et al.}}

\DeclareMathOperator{\ii}{i}

\newcommand{\Cmax}{C_{\rm max}}

\begin{document}

\title{
Landau levels of Majorana fermions in a spin liquid
}

\author{Stephan Rachel}
\affiliation{Institut f\"ur Theoretische Physik, Technische Universit\"at Dresden,
01062 Dresden, Germany}
\author{Lars Fritz}
\affiliation{Institute for Theoretical Physics and Center for Extreme Matter and Emergent Phenomena,
Utrecht University, Leuvenlaan 4, 3584 CE Utrecht, The Netherlands}
\author{Matthias Vojta}
\affiliation{Institut f\"ur Theoretische Physik, Technische Universit\"at Dresden,
01062 Dresden, Germany}

\date{\today}


\begin{abstract}
Majorana fermions, originally proposed as elementary particles acting as their own antiparticles, can be realized in condensed-matter systems as emergent quasiparticles, a situation often accompanied by topological order.
Here we propose a physical system which realizes Landau levels -- highly degenerate single-particle states usually resulting from an orbital magnetic field acting on charged particles -- for Majorana fermions. This is achieved in a variant of a quantum spin system due to Kitaev which is distorted by triaxial strain. This strained Kitaev model displays a spin-liquid phase with charge-neutral Majorana-fermion excitations whose spectrum corresponds to that of Landau levels, here arising from a tailored pseudo-magnetic field. We show that measuring the dynamic spin susceptibility reveals the Landau-level structure by a remarkable mechanism of probe-induced bound-state formation.
\end{abstract}
\maketitle


In 1937 E. Majorana suggested \cite{majorana} that a real wavefunction can describe spin-$1/2$ quantum particles that are their own antiparticles. Consequently, these particles are charge-neutral and decouple from the electromagnetic field, making them hard to detect. While no observation of Majorana fermions as elementary particles has been reported to date, their possible realization as emergent quasiparticles in condensed-matter systems has attracted enormous attention.
Specifically, it was realized that superconductors provide a natural habitat, where Bogoliubov quasiparticles at zero energy have properties of Majorana fermions \cite{beenakker_rev}. Moreover, a seminal work of Kitaev \cite{kitaev06} demonstrated that dispersive Majorana fermions can emerge as effective degrees of freedom in spin-liquid phases of frustrated quantum magnets \cite{balents10}. This has triggered an intense search for materials which come close to realizing Kitaev's spin-liquid model on the honeycomb lattice, with {\aio} (A=Na,Li) (Refs.~\onlinecite{jackeli09,Cha10,gegenwart12,kee14,chun15}) and $\alpha$-RuCl$_3$ (Ref.~\onlinecite{banerjee15}) currently being the best candidates.

Here we show that the properties of emergent Majorana fermions in quantum magnets can be engineered by controlled lattice distortions. We use spatially inhomogeneous exchange couplings, generated by applying a suitable strain pattern, to transform the linear Dirac-like dispersion of low-energy Majorana fermions in the Kitaev model into a sequence of pseudo-Landau levels. The idea of strain-induced Landau levels was in fact first theoretically proposed \cite{guinea09} and then subsequently realized \cite{levy10} for electrons in two-dimensional graphene. Our work adapts this concept to the charge-neutral fractionalized quasiparticles of a topological spin liquid. We compute the dynamic spin susceptibility which turns out to display sharp excitations at isolated energies reflecting the Landau-level structure of the strained spin liquid.

\begin{figure}[t!]
\centering
\includegraphics[scale=0.7]{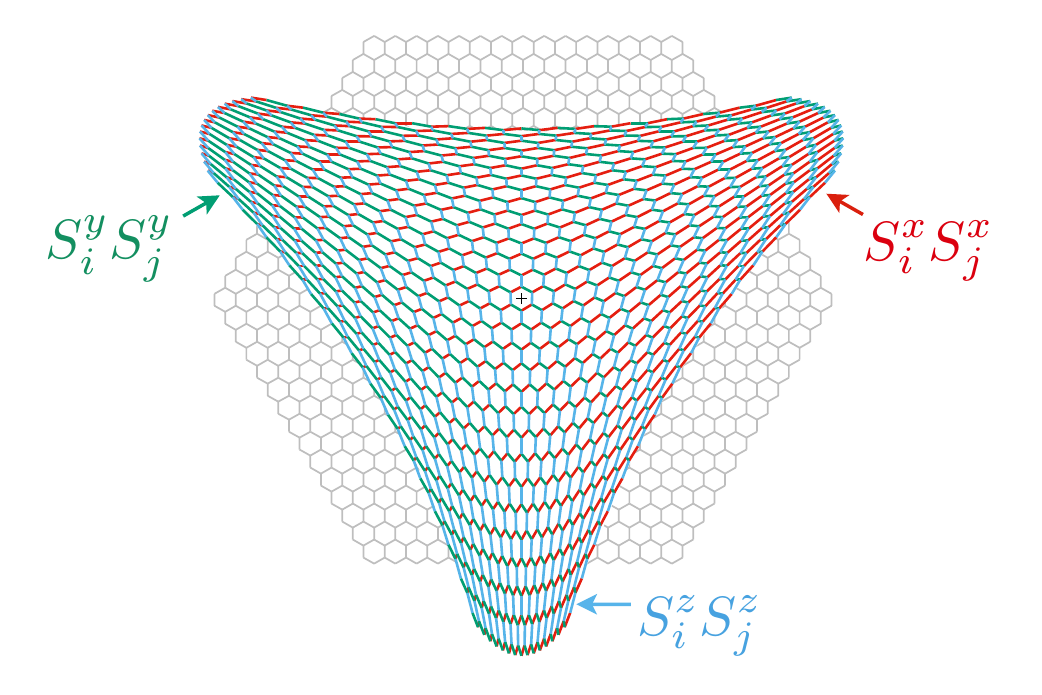}
\caption{
Lattice setup: Honeycomb-lattice flake with $N=675$ unit cells where bonds with different color correspond to Ising interactions $J^\alpha$ of different spin components $\alpha$.
The distorted lattice visualizes the displacement according to Eq.~\eqref{displace} arising from triaxial strain; the unstrained hexagonal flake is shown in the back in grey. Longer (shorter) bonds correspond to weaker (stronger) exchange couplings $J_{ij}^\alpha$.
}
\label{fig:setup}
\end{figure}

\textit{Model.}
The Kitaev model \cite{kitaev06} describes quantum spins $1/2$ on a honeycomb lattice subject to spin-anisotropic Ising (or ``compass'') interactions. The Hamiltonian, generalized to spatially varying couplings, reads
\begin{equation}
\label{hk}
\HK =
-\sum_{\langle ij\rangle_x} J_{ij}^x \hat{\sigma}_i^x \hat{\sigma}_j^x
-\sum_{\langle ij\rangle_y} J_{ij}^y \hat{\sigma}_i^y \hat{\sigma}_j^y
-\sum_{\langle ij\rangle_z} J_{ij}^z \hat{\sigma}_i^z \hat{\sigma}_j^z
\end{equation}
where $\hat{\sigma}^{\alpha}$ are Pauli matrices, and $\langle ij \rangle_\alpha$ denotes an $\alpha$ bond as shown in Fig.\,\ref{fig:setup}, with $\alpha=x,y,z$.
This model displays an infinite set of constants of motion, which can be interpreted as $\mathbb{Z}_2$ fluxes through the closed loops of the lattice. Upon representing each spin $\hat{\sigma}$ by four Majorana fermions $\hat{b}^x$, $\hat{b}^y$, $\hat{b}^z$ and $\hat{c}$, with $\hat{\sigma}_i^\alpha=\ii \hat{b}_i^\alpha \hat{c}_i$, the model \eqref{hk} takes the form
\begin{equation}
\label{hm}
\HMh = \ii \sum_{\langle ij\rangle}J^\alpha_{{ij}} \hat{u}_{ij}\hat{c}_i \hat{c}_j,
\end{equation}
where $\hat{u}_{ij}\equiv \ii\hat{b}_i^{\alpha_{ij}}\hat{b}_j^{\alpha_{ij}}$ and $\hat{u}_{ij}=-\hat{u}_{ji}$. The operators $\hat{u}_{ij}$ represent conserved quantities with eigenvalues of $u_{ij}=\pm 1$ which determine the values of the $\mathbb{Z}_2$ fluxes. Hence the Hamiltonian $\HMh$ \eqref{hm} represents a nearest-neighbor hopping problem for the $c$ (or ``matter'') Majorana fermions which are coupled to a static $\mathbb{Z}_2$ gauge field. Its exact solution can be written in terms of canonical-fermion excitation modes with non-negative energies $\e_m$. For homogeneous and isotropic couplings, $J_{ij}^\alpha\equiv J$, it describes a $\mathbb{Z}_2$ spin liquid with matter-fermion excitations displaying a gapless Dirac spectrum.

For our case of inhomogeneous couplings we shall solve the Kitaev model numerically on finite-size lattices. To this end, Eq.~\eqref{hm} is re-written into the following bilinear Hamiltonian for the matter Majorana fermions
\begin{equation}
\label{hmflux}
\HM = \frac{\ii}{2}\left(\hat{c}^T_A \, \hat{c}^T_B\right) \begin{pmatrix}
0 & M \\
-M^T & 0
\end{pmatrix}
\begin{pmatrix}
\hat{c}_A \\
\hat{c}_B
\end{pmatrix}
\end{equation}
where $M$ is an $N \times N$ matrix with elements $M_{ij}=J^\alpha_{ij} u_{ij}$ and $\hat{c}_{A(B)}$ is a vector of length $N$ of Majorana operators on the $A(B)$ sublattice, and $N$ the number of unit cells. We construct the matrix $M$ for a given set of couplings $\{J^\alpha_{ij}\}$ and diagonalize the problem via a singular-value decomposition which yields the excitation energies $\e_m$ and the eigenmodes. From these we calculate the local density of states (LDOS) as the local spectral density of the $c$ fermions, for details see supplement \cite{supp}.

\textit{Majorana Landau levels.}
In the context of graphene, it has been theoretically shown \cite{guinea09,voz_rev10} that a spatial modulation of hopping energies mimics the effect of a vector potential in Dirac fermion systems. Near the Dirac energy, this emergent vector potential can be expressed through the strain tensor $u_{ij}$ as $\vec A \propto \pm ( u_{xx} - u_{yy} , -2 u_{xy} )^T$, with opposite sign for electrons belonging to the two Dirac cones (or ``valleys'') located at $\vec K=(4\pi/(3\sqrt{3}),0)$ and $\vec K'=(-4\pi/(3\sqrt{3}),0)$ (with the lattice constant $a_0$ set to unity), such that time-reversal symmetry is preserved. If the resulting pseudo-magnetic field, $\vec B={\rm rot}\vec A$, is sufficiently homogeneous -- applying e.g. to triaxial strain patterns as shown in Fig.~\ref{fig:setup} -- it can induce single-particle pseudo-Landau levels very similar to Landau levels in a physical magnetic field. The system then displays a so-called quantum valley Hall effect\,\cite{guinea09}, \ie it combines a chiral and an antichiral quantum Hall effect at the valleys $\vec K$ and $\vec K'$, respectively.
For triaxial strain the displacement vector is given by \cite{guinea09,peeters13}
\begin{equation}
\label{displace}
\vec{U}(x,y) = \bar{C} \big(2xy, x^2-y^2 \big)
\end{equation}
where $\bar{C}$ (measured in units of $1/a_0$) parameterizes the distortion, and $u_{ij} = (\partial_i U_j + \partial_j U_i)/2$. This displacement yields -- to first order in $\bar{C}$ -- a homogeneous pseudo-magnetic field whose strength is proportional to $\bar{C}$.

Adapting the idea of strain-induced artificial gauge fields to spin systems, we study the Kitaev model \eqref{hk} with spatially modulated exchange couplings.
Given that the Majorana-fermion hopping matrix elements in Eq.\,\eqref{hm} are given by the exchange couplings $J^{\alpha}_{ij}$, we choose
\begin{equation}
\label{ourjij}
J^{\alpha}_{ij} = J^\alpha \left[ 1- \beta (|\vec{\delta}_{ij}|/a_0 - 1)\right]
\end{equation}
where we calculate the distance $\vec{\delta}_{ij}=\vec{R}_i+\vec{U}_i-\vec{R}_j-\vec{U}_j$ using the $\vec{U}(x,y)$ from Eq.\,\eqref{displace} evaluated at the lattice positions $\vec{R}_i$ of the undistorted honeycomb lattice. The factor $\beta$ encodes the strength of magnetoelastic coupling, and the parameter (dubbed ``strain'' below)
$C = \bar{C}\beta$
will enter our simulations as a measure of the modulations of the $J^{\alpha}_{ij}$.
Throughout the paper we choose $\beta=1$; larger values of $\beta$ (hence smaller $\bar{C}$) reduce non-linearities in the strain pattern but are less realistic for transition-metal oxides.
We note that, for a real material, Eq.~\eqref{ourjij} represents a linear approximation to the full dependence of the exchange constant on the bond length \cite{peeters13}.

\begin{figure*}
\centering
\includegraphics[scale=0.51]{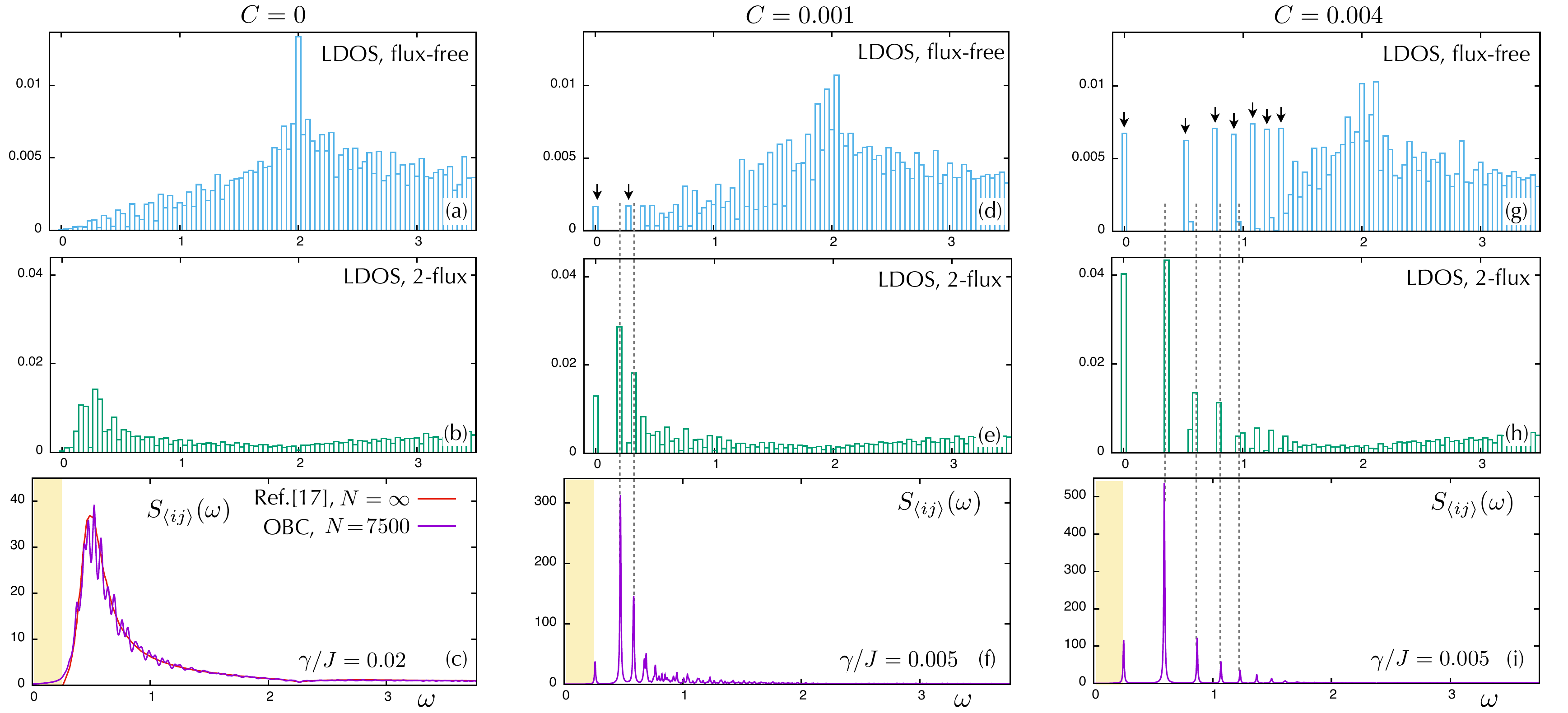}
\caption{Simulation results:
Histogram of the matter-fermion LDOS in the flux-free (top, blue) and two-flux sectors (middle, green), together with the dynamical spin correlator $S_{\langle ij \rangle}(\w)$ with $S_{\langle ij \rangle}=S_{ii}+S_{jj}+2S_{ij}$ (bottom, purple), for the cases
(a-c): unstrained ($C=0$),
(d-f): moderate strain ($C=1\cdot 10^{-3}$), and
(g-i): large strain ($C=4\cdot 10^{-3}$).
All quantities are measured in the center of a sample with $2N=15\,000$ sites. The $S(\w)$ plots employ a Lorentzian broadening of width $\gamma$; moreover the energy axis has been shifted with respect to the LDOS plots by the amount of the local flux gap\cite{supp} $\Delta_{ij}$ (yellow region). Panel (c) also shows $S(\w)$ for the infinite homogeneous system \cite{knolle}.
While the LDOS of the flux-free sector [panel (d) and (g)] nicely shows the emergence of Majorana Landau levels (black arrows), the two-flux sector features bound states in the gaps between the Landau levels [panels (e) and (h)] -- these are detected in $S_{\langle ij \rangle}(\w)$ as sharp peaks (dashed lines are guide-to-the-eye).
}
\label{fig:res1}
\end{figure*}

The inhomogeneous Kitaev model \eqref{hk} with couplings given by Eq.\,\eqref{ourjij} is expected to display a $\mathbb{Z}_2$ spin-liquid phase with an extremely unusual excitation spectrum: While the $ \mathbb{Z}_2$ fluxes remain static, the matter Majorana fermions will form highly degenerate Landau levels.
Given that a physical magnetic applied applied to a Kitaev magnet has an entirely different effect -- it induces flux dynamics and leads to a gapped matter spectrum \cite{kitaev06} -- strain represents a unique way to generate Majorana Landau levels.

\begin{figure}[b!]
\centering
\includegraphics[scale=0.7]{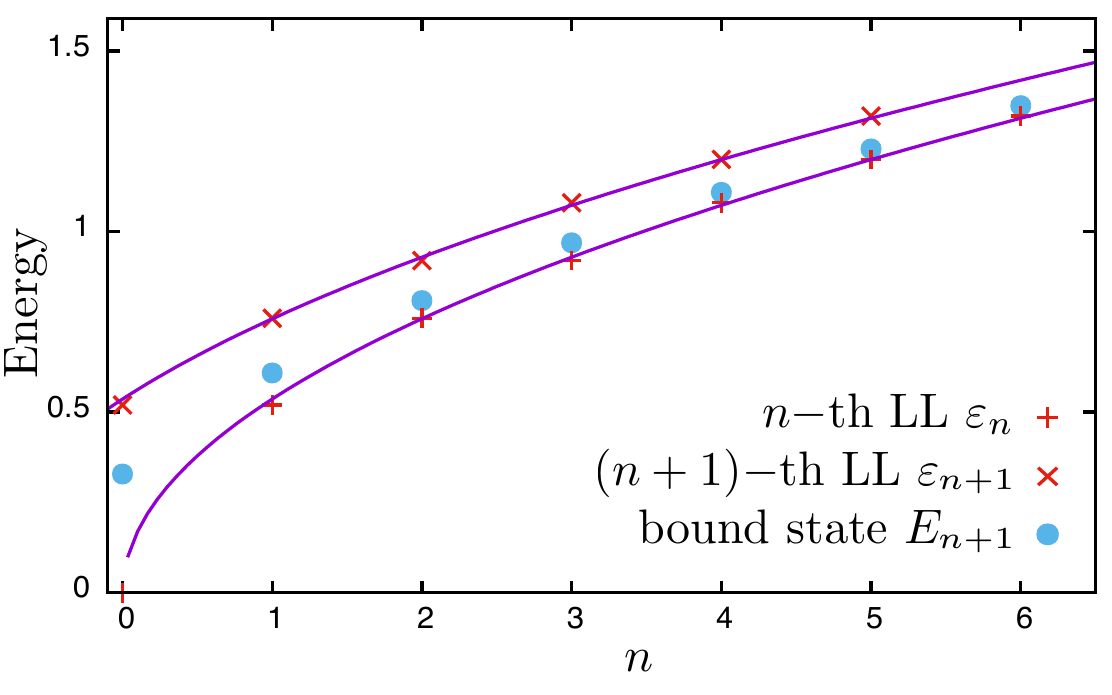}
\caption{
Energy positions of the Landau levels $\ELL_n$ as observed in the matter-fermion LDOS of the flux-free sector, shown together with the bound-state energies $\EB_n$ in the two-flux sector; the $\EB_n$ are identical to the peak positions in the correlator $S_{\langle ij \rangle}(\w)$ with the local flux gap $\Delta_{ij}$ subtracted [except for the lowest peak which originates from the LLL; hence our numbering for $\EB_n$ starts with $n=1$].
The data correspond to that of Fig.~\ref{fig:res1}(g,i) with $C=0.004$.
The lines are fits of $\ELL_n$ to a $\sqrt{n}$ dependence; both $\ELL_n$ and $\ELL_{n+1}$ are shown for visualization purposes such that the Landau-level gap is visible for each $n$.
}
\label{fig:bsenergies}
\end{figure}

This is well borne out by our numerical results, obtained for finite-size Kitaev systems of hexagonal shape, Fig.~\ref{fig:setup}, with open zigzag boundaries and lattice sizes up to $2N=15\,000$ sites. We focus on the case of isotropic couplings and choose $J^\alpha=J$ as energy unit. We limit the strain $C$ to be smaller than $\Cmax$, the latter being defined as the largest $C$ where all $J_{ij}>0$ \cite{supp}. For $C<\Cmax$ we find that the ground state is located in the flux-free sector where all $u_{ij}$ can be chosen to be $+1$.
We monitor the LDOS of the matter fermions \cite{supp} which is shown in the top panels of Fig.~\ref{fig:res1}: In the unstrained case $C=0$ this is the familiar honeycomb-lattice DOS  [panel (a)], with $\rho(\w)\propto\w$ at low energies, with the difference to graphene that only half of the spectrum corresponding to $\w\geq0$ is realized due to the Majorana nature of the matter fermions.
Importantly, the results for finite strain show clear Landau-level peaks at low energies highlighted by the black arrows in panels (d) and (g). In particular, the Landau-level energies $\ELL_n$ display the scaling $\ELL_n\propto \sqrt{n C}$ with the lowest Landau level (LLL) located at zero energy, Fig.~\ref{fig:bsenergies}, characteristic of honeycomb-lattice Dirac fermions \cite{giesbers}.

\textit{Dynamic structure factor.}
In order to detect the Landau-level structure, we propose to measure dynamic spin correlations
\begin{equation}
\label{sfac}
S^{\alpha \beta}_{ij}(t)=\langle\hat{\sigma}_i^\alpha(t)\hat{\sigma}_j^\beta(0)\rangle
\end{equation}
whose Fourier transform is -- in a magnet -- accessible by neutron scattering experiments. We restrict our attention to zero temperature where $S^{\alpha \beta}_{ij}(\w)$ is proportional to the imaginary part of the dynamic spin susceptibility.
The evaluation of the dynamic spin correlations in the Kitaev model has been first discussed in Ref.~\onlinecite{knolle}. Specifically, the application of the operator $\hat{\sigma}_i^\alpha = \ii \hat{b}_i^\alpha \hat{c}_i$ changes the $\mathbb{Z}_2$ fluxes in the plaquettes adjacent to the $\alpha$ bond emanating from site $i$. Starting from the ground state $\ket{0}$ which is flux-free, the intermediate state $\ket{\lambda}$ is then a state with a locally excited flux pair, i.e.,  the hopping problem of matter fermions in the intermediate state involves a local ``flux impurity'' $\hat{V} = -2\ii J^{\alpha}_{ij} \hat{c}_i \hat{c}_j$. The fact that fluxes are static implies that all correlators beyond nearest-neighbor sites are strictly zero.

The numerical calculation of a zero-temperature dynamic structure factor $S_{ij}(\w)$ involves two diagonalizations, one in the flux-free sector with all $u=1$ and one in the excited-state flux sector with all $u=1$ except $u_{ij}=-1$ on the bond involving the measured site(s). The eigenvectors in both flux sectors are then used to construct the matrix elements entering $S_{ij}(\w)$. We employ the single-mode approximation \cite{knolle}, with details given in the supplement \cite{supp}.

For the structure factor of the strained Kitaev model, a central observation is that the impurity $\hat{V}$ induces a sequence of single-particle bound states which energetically lie between the Landau levels. This can be nicely seen in panels (e) and (h) of Fig.~\ref{fig:res1} which show the LDOS in the two-flux sector, measured on the flipped bond. This LDOS displays essentially the same Landau levels as the corresponding flux-free system, but shows additional sharp peaks of even higher weight at energies $\EB_n$ in between the Landau levels -- these peaks arise from isolated states localized near the flipped bond.
We note that related bound states have been reported recently in a study of spin excitations in the field-induced non-Abelian phase of the Kitaev model \cite{knolle15}. More broadly, we recall that in-gap bound states occur frequently in one- and two-dimensional systems because the real part of the single-particle Green's function diverges at gap edges. In fact, recent theory work on topological band insulators has suggested that the {\em generic} occurrence of impurity-induced bound states in a two-dimensional gapped system is a signature of its topological nature \cite{zaanen15}.
For the present case of bond impurities in the triaxially strained honeycomb lattice, we have checked that bound states between the Landau levels occur indeed generically, i.e., independent of the impurity strength \cite{supp}, hinting at the topological character of the matter-fermion sector.

Remarkably, these Majarona-fermion bound states enable a unique detection of the Landau-level structure: The largest matrix elements for the correlator $S_{\langle ij \rangle}(\w)$ arise from these bound states in the two-flux sector, such that $S_{\langle ij \rangle}(\w)$ displays sharp peaks at $\EB_n + \Delta_{ij}$ where $\Delta_{ij}$ is the local flux gap \cite{zschocke15,supp}. As can be seen in the lower panels of Fig.\,\ref{fig:res1}, these bound-state peaks dominate over the peaks arising from the Landau levels themselves.

In Fig.~\ref{fig:bsenergies} we show the energetic positions of the bound-state peaks in $S_{\langle ij \rangle}(\w)$, with the local flux gap $\Delta_{ij}$ subtracted, and compare them to the energies of the Landau levels found in the matter-fermion LDOS of the flux-free system. The plot shows that the bound-state energies track the energy dependence $\ELL_n \propto \sqrt{n C}$ of the Landau levels, approaching the lower edge of the respective gap with increasing $n$.
Given that sharp bound states can only arise in gapped systems, the observation of a series peaks in $S_{\langle ij \rangle}(\w)$ with a $\sqrt{n}$ energy dependence (after subtracting a constant corresponding to $\Delta_{ij}$) represents a direct signature of the peculiar excitation spectrum -- Landau levels separated by gaps -- of the strained spin liquid.


\textit{Discussion.}
Our results raise a number of further issues.
First, we note that the simultaneous application of strain and of a physical magnetic field has non-trivial effects, which can be analyzed perturbatively by projecting onto the flux-free sector \cite{kitaev06}. We have found that the LLL of the strained Kitaev model is shifted to a finite energy, while otherwise the spectrum of the matter Majorana fermions remains qualitatively unchanged. The excitations of the now fully gapped system obey non-Abelian statistics similar to the elementary excitations of Kitaev's B-phase\,\cite{kitaev06} subject to a magnetic field.
%
Second, and perhaps most interesting, is the fate of the Landau-level spin liquid when interactions beyond the Kitaev model are included \cite{rosch15}. While this question is beyond the scope of this work, it is likely that the large degeneracy of the many-body ground state will be lifted in favor of potentially exotic states driven by interactions between the Majorana fermions, akin to the physics of the fractional quantum Hall effect.

Realizations of our system appear possible using strained thin films of honeycomb-lattice iridates, where it has been proposed that homogeneous pressure or strain may drive the material into a Kitaev spin-liquid regime. Landau levels can then be generated using shapes resulting from triaxial strain, circular arcs \cite{strain-arc}, or nanobubbles \cite{levy10}. It has to be kept in mind, however, that inhomogeneous strain may induce additional interactions as it deforms the nearly 90-degree Ir-O-Ir bonds.
A second possibility is offered by advances in designing artificial molecular structures on surfaces, where the electronic structure of strained graphene has already been demonstrated \cite{manoharan12}.
A third option is by simulating the Hamiltonian \eqref{hk} using ultracold atomic gases. Phase plates \cite{phaseplate} or digital mirror devices \cite{DMD} can be used to directly project the distorted lattice onto a 2D cloud of atoms, and proposals for realizing artificial spin-orbit coupling have been made \cite{cold_soc}.

In summary, we have demonstrated how to engineer a novel state of matter, where the emergent degrees of freedom of a fractionalized spin liquid -- charge-neutral Majorana fermions -- display a Landau-level structure of excitation energies. We have shown that this Landau-level structure can be efficiently detected in measurements of dynamic spin correlations, thanks to a mechanism of probe-induced bound-state formation.
%


We thank J.\ Knolle, Y.\ M.\ Lu, T.\ Meng, M.\ Neek-Amal, F.\ Peeters, and F.\ Zschocke for discussions.
This research was supported by the DFG through SFB 1143 and GRK 1621
as well as by the Helmholtz association through VI-521.


\end{document}


\title{Supplemental material:\\
Landau levels of Majorana fermions in a spin liquid}
\author{Stephan Rachel}
\affiliation{Institut f\"ur Theoretische Physik, Technische Universit\"at Dresden, 01062 Dresden, Germany}
\author{Lars Fritz}
\affiliation{Institute for Theoretical Physics and Center for Extreme Matter and Emergent Phenomena, Utrecht University, Leuvenlaan 4, 3584 CE Utrecht, The Netherlands}
\author{Matthias Vojta}
\affiliation{Institut f\"ur Theoretische Physik, Technische Universit\"at Dresden, 01062 Dresden, Germany}


\date{\today}

\maketitle

\section{Exact solution of the Kitaev model}

The spin-liquid model of Kitaev \cite{kitaev06} describes spins $1/2$ on a honeycomb lattice which interact via bond-dependent nearest-neighbor Ising spin exchange,
\begin{equation}\label{ham:spin}
\mathcal{H}_{\rm K} =
-\sum_{\langle ij \rangle_x} J_{ij}^x \hat\sigma_i^x \hat\sigma_j^x
-\sum_{\langle ij \rangle_y} J_{ij}^y \hat\sigma_i^y \hat\sigma_j^y
-\sum_{\langle ij \rangle_z} J_{ij}^z \hat\sigma_i^z \hat\sigma_j^z
\end{equation}
with $\hat\sigma_i^\alpha$ being Pauli matrices. For the homogeneous isotropic case, $J_{ij}^\alpha\equiv J$, the Kitaev model possesses a discrete $\mathbb{Z}_3$ symmetry of a combination of 120 degree real-space and simultaneous spin rotations of the form $\hat\sigma^x \to \hat\sigma^y$, $\hat\sigma^y\to\hat\sigma^z$, and $\hat\sigma^z\to\hat\sigma^x$. As usual, the two sublattices of the honeycomb lattice will be labelled A and B.

Following Kitaev \cite{kitaev06}  one rewrites the spin Hamiltonian \eqref{ham:spin} in terms of four Majorana fermions per site, $\hat b^x$, $\hat b^y$, $\hat b^z$, and $\hat c$, with $\hat\sigma_i^\alpha = \ii \hat b_i^\alpha \hat c_i$. In the new operators, the Hamiltonian \eqref{ham:spin} becomes
\begin{equation}\label{ham:majo1}
\mathcal{H}_{\hat u} = \ii \sum_{\langle ij \rangle_\alpha} J_{ij}^\alpha \, \hat u_{ij} \, \hat c_i \, \hat c_j
\end{equation}
with the ``bond operator'' $\hat u_{ij} \equiv \ii \hat b_i^\alpha \hat b_j^\alpha$ which is antisymmetric under exchange of $i$ and $j$ (and $i$ taken from the A sublattice). The problem can be solved by noting that all $\hat u_{ij}$ are constants of motion: they commute with the Hamiltonian $\mathcal{H}_{\hat u}$ and with each other, their eigenvalues being $u_{ij}=\pm 1$. Every particular choice of a given set $\{ u_{ij} \}$ reduces the Hamiltonian \eqref{ham:majo1} into a billinear which is readily solved,
\begin{equation}\label{ham:majo2}
\mathcal{H}_u = \frac{\ii}{2} \Big( \hat c_A^T~\hat c_B^T \Big) \left( \begin{array}{cc} 0 & M \\[5pt] -M^T & 0 \end{array}\right) \left(\begin{array}{c} \hat c_A \\[5pt] \hat c_B \end{array}\right)\ .
\end{equation}
For a honeycomb lattice with $2N$ lattice sites, $M$ is a real $N\times N$ matrix with elements $M_{jk}=J^{\alpha}_{j k} u_{jk}$ and the vector $\hat c_{A(B)}$ contains $N$ matter Majorana operators of the A (B) sublattice. Using a singular value decomposition one finds $M=USV^T$ with orthogonal $N\times N$ matrices $U$ and $V$ and the positive semi-definite diagonal matrix $S$ with elements $\epsilon_i$, $i=1,\ldots, N$. Denoting the Majorana fermions on sublattice A (B) in unit cell $\bs{r}$ as $\hat c_{A(B),\bs{r}}$, the Majorana eigenmodes are given by
\begin{equation}\label{Eq4}
\hat c'_{A,m}=\sum_{\bs{r}} U^T_{m,\bs{r}} \hat c_{A,\bs{r}} \quad\hbox{and}\quad
\hat c'_{B,m}=\sum_{\bs{r}} V^T_{m,\bs{r}} \hat c_{B,\bs{r}}
\end{equation}
which can be combined into complex fermions
\begin{equation}
\hat a_m^{\phantom{\dag}} = \frac{1}{2}\left( \hat c'_{A,m} + \ii \hat c'_{B,m} \right)\ .
\label{adef}
\end{equation}
Using the matrix identity
\begin{equation}\label{bloch-matrix}
\left(\!\begin{array}{cc} 0& USV^T \\[5pt] -VS U^T & 0\end{array}\!\right) =
\left(\!\begin{array}{cc}U&0\\[5pt] 0&V\end{array}\right)\left(\!\begin{array}{cc}0&S\\[5pt] -S&0\end{array}\!\right)\left(\!\begin{array}{cc}U^T & 0\\[5pt] 0 & V^T \end{array}\!\right)
\end{equation}
we can rewrite the Hamiltonian \eqref{ham:majo2},
\begin{eqnarray}
\nn \mathcal{H}_u &=& \frac{\ii}{2} \Big( \hat c_A^T ~ \hat c_B^T \Big)  \left(\!\begin{array}{cc}U&0\\[5pt] 0&V\end{array}\!\right)\left(\!\!\begin{array}{cc}0&S\\[5pt] -S&0\end{array}\!\right)\left(\!\begin{array}{cc}U^T & 0\\[5pt] 0 & V^T \end{array}\!\!\!\right) \left(\!\!\begin{array}{c}\hat c_A\\[5pt] \hat c_B\end{array}\!\!\right)\\[5pt]
\nn &=&\frac{\ii}{2} \left( {\hat c'_A}\!\!\,^T ~ {\hat c'_B}\!\!\,^T \right) \left(\begin{array}{cc}0&S\\[5pt] -S & 0 \end{array}\!\right)
\left(\begin{array}{c}\hat c'_A\\[5pt] \hat c'_B\end{array}\right)\\[5pt]
&=& \ii \sum_{m=1}^N \epsilon_m \hat c'_{A,m} \hat c'_{B,m}  = \sum_{m=1}^N \epsilon_m \left( 2 \hat a_m^\dag \hat a_m^\pd -1 \right) \label{ham:majo3}\,.
\end{eqnarray}
The matter-fermion excitation energies are thus given by the singular values $\epsilon_m$ of the matrix $M$; the ground-state energy for a given choice of $\{u_{ij}\}$ is $E_{0,u}=-\sum_m \epsilon_m$. Following Kitaev we will refer from now on to the choices $\{u_{ij}\}$ as flux sectors. The case where all $u_{ij}=+1$ is referred to as the {\it flux-free} sector. The case where all $u_{ij}=+1$ except a single $u_{\mu\nu}=-1$ is referred to as a {\it single-flux sector} if the bond $\langle \mu\nu\rangle$ is located at the boundary of the sample and as {\it two-flux sector} otherwise.

To monitor the eigenmodes of the matter Majorana fermions, we can calculate their (global) density of states (DOS), $\rho(\omega) = \sum_n \delta(\omega- 2\epsilon_n )$. To resolve spatial structure, we also consider the corresponding local density of states (LDOS) which is given by
\begin{equation}
\rho_j(\omega) = \frac{1}{2} \sum_n  \, U_{nj}^2 \, \delta (\omega - 2\epsilon_n)
\end{equation}
for a site $j$ on the A sublattice; for a site $j$ on the B sublattice, $U_{nj}$ is replaced by $V_{nj}$ where $U_{nj}$ ($V_{nj}$) are the matrix elements of $U$ ($V$).

\begin{figure}[h!]
\centering
\includegraphics[scale=0.6]{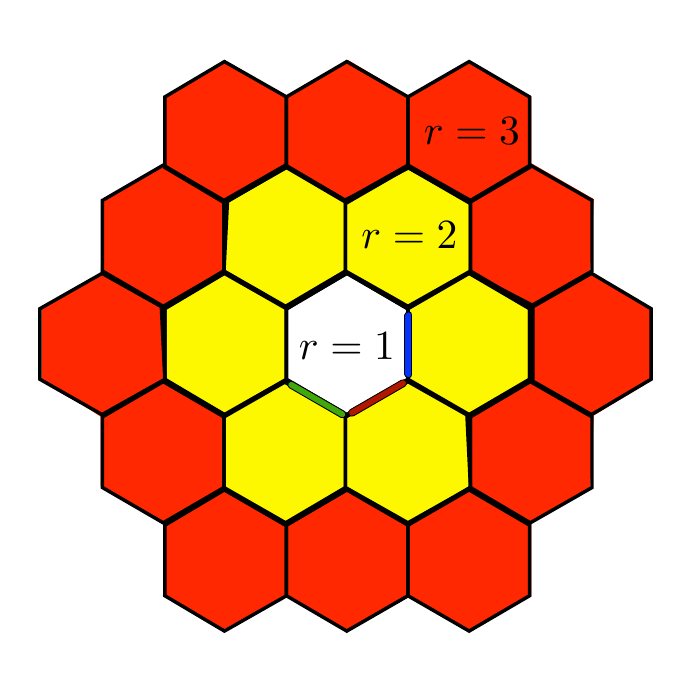}
\caption{Illustration of a hexagonal shaped honeycomb flake consisting of $r$ ``rings''.
Shown are $r=1$ (white), $r=2$ (yellow + white), $r=3$ (red + yellow + white). The number of unit cells is $N=3r^2$ corresponding to $2N$ spins.
}
\label{fig:hc-flake}
\end{figure}

In our numerical calculations we consider hexagonal flakes with open boundaries (see Fig.\,\ref{fig:hc-flake}), consisting of $r$ rings, with the number of unit cells being $N=3r^2$. Most runs are performed for $r=50$, yielding $2N=15\,000$ spins.

\section{Dynamical spin correlations}

The dynamical spin correlation function, Eq.~{\eqsoft} of the main paper, can be expressed in terms of Majorana fermions and yield the following Lehmann representation at zero temperature\,\cite{knolle,zschocke15}, here written separately for the on-site correlator
\begin{equation}
\label{sii}
S^{\alpha\beta}_{ii}(\w) =  2\pi \sum_\lambda |\bra{M_0}\hat{c}_i\ket{\lambda}|^2 \delta(\w-(E_\lambda-E_0)) \delta_{\alpha \beta}
\end{equation}
and the off-site correlators
\begin{align}
\label{sij}
S^{\alpha \beta}_{ij}(\w) = 2\pi F_{ij}^\alpha \sum_\lambda &\bra{M_0}\hat{c}_i\ket{\lambda} \bra{\lambda}\hat{c}_j\ket{M_0} \\ &\times \delta(\w-(E_\lambda-E_0)) \delta_{\alpha \beta} \delta_{\langle ij \rangle_\alpha},
\nonumber
\end{align}
where $\delta_{\langle ij \rangle_\alpha}$ is non-zero only if $i$ and $j$ are nearest neighbors connected by an $\alpha$ bond, i.e., $S_{ij}$ vanishes beyond nearest neighbors. In both cases $\ket{M_0}$ is the matter-fermion ground state in the ground-state flux sector, assumed to be flux-free, and $\sum_{\lambda}$ runs over all matter-fermion states in the two-flux sector. $E_0$ and $E_\lambda$ are the corresponding many-body energies. The prefactor $F_{ij}^\alpha=\{-1,\ii,-\ii\}$ depending on the spin component.
Note that we can safely ignore the fermion parity condition \cite{loss,zschocke15} due to open boundaries, such that $\ket{M_0}$ has no excited matter fermions. For the intermediate state $\ket{\lambda}$ we will restrict ourselves to states with a single matter fermion (single-mode approximation); these states contribute more than 97\% of the spectral weight in the homogeneous case \cite{knolle}. With strain applied, we can expect this approximation to be even better, because the main weight is contained in the bound-state peaks whose contributions are captured exactly in the single-mode approximation.

In the homogeneous isotropic Kitaev model, the momentum-dependent dynamic spin structure factor $S^{\alpha\alpha}(\vec{q},\w) = (1/N) \sum_{ij} S^{\alpha\alpha}_{ij}(\w) e^{-\ii\vec{q}\cdot\vec{r}_{ij}}$ -- accessible by neutron scattering -- displays a continuum of spectral weight above the energy of the flux gap, $\Delta=0.26J$, with little momentum dependence \cite{knolle}. This reflects the fact that a spin flip decays into a matter and a flux excitation, the latter having a momentum-independent energy.


For our inhomogeneous Kitaev model, the computation of the spin correlator proceeds in the following steps:
\begin{enumerate}
\item We construct the matrix $M$ in \eqref{bloch-matrix} for the flux-free sector (\ie all $u_{ij}=+1$). This matrix is real and of dimension $N\times N$, its entries require to specify the values of $J^\alpha$ and the strain $C$.
\item We perform a singular value decomposition (SVD) into $M=USV^T$ where $U$ and $V$ contain the normalized singular vectors and are orthogonal, $U^T=U^{-1}$ and $V^T=V^{-1}$. The diagonal matrix $S$ contains the singular values -- these are identical to the non-negative part of the spectrum of an equivalent hopping problem of {\em canonical} fermions with hopping energies $M_{jk}$.
\item We obtain the $M'$ matrix of a particular two-flux sector, defined by a single sign-flipped bond $\langle ij\rangle$ (equivalent to multiplying a single matrix element in $M$ by $-1$). Now we also perform the SVD for the matrix $M'$ and obtain corresponding $U'$ and $V'$ matrices as well as the singular values $\epsilon^{(2)}_m$. We denote the matter-fermion ground state in this two-flux sector by $\ket{\lambda_0}$ and the canonical fermions describing the corresponding excitations [analogous to those in Eq.~\eqref{adef}] by $\hat{b}_m$.
\item We define the matrix
\begin{equation}
X=\frac{1}{2}\Big( {U'}^T U + {V'}^T V \Big)
\end{equation}
and compute its determinant and its inverse, $\det{X}$ and $X^{-1}$, respectively.
\item For an intermediate state with a single matter-fermion excitation, $\ket{\lambda_m} = \hat{b}_m^\dagger\ket{\lambda_0}$, matrix elements for $\hat{c}$ operators on the A and B sublattices are obtained as
\begin{eqnarray}
\label{def:RA}
\bra{M_0} \hat c_{A,i} \ket{\lambda_m} &=&  \sqrt{|\det X|} \left( U X^{-1} \right)_{i,m}\,,\\[5pt]
\label{def:RB}
\bra{\lambda_m} \hat c_{B,j} \ket{M_0} &=& \ii  \sqrt{|\det X|} \left( V X^{-1} \right)_{j,m}\,.
\end{eqnarray}
\item Eventually we calculate the correlators \eqref{sii} and \eqref{sij},
with the ground state energy of the flux-free sector $E_0$
and with the one-particle energies of the considered two-flux sector, with
\begin{equation}
{E_\lambda}_m = E_0^{(2)} + 2 \epsilon^{(2)}_m = -\sum_{m'=1}^{N} \epsilon_{m'}^{(2)} +  2 \epsilon_m^{(2)}\ .
\end{equation}

\end{enumerate}


\section{Finite-size and boundary effects}

The displacement pattern as described by Eq.~{\eqdisplace} is incompatible with periodic boundary conditions, such that open boundaries need to be employed. Then, the pseudo-magnetic field in a finite-size system as in Fig.~{\figsetup} is not perfectly homogeneous. In addition, low-energy edge states occur.

\subsection{Maximal strain}

The assumed linear dependence of the exchange couplings on bond lengths, Eq.~{\eqinhomj}, implies that there exists a maximal value of the strain, $\Cmax$, such that for $C>\Cmax$ exchange couplings of the boundary of the sample become negative. The value of $\Cmax$ decreases with system size; concrete values are $\Cmax = 0.021$ for $2N=1350$ and $\Cmax = 0.0062$ for $2N=15000$. All results shown in this paper are for strain values $C<\Cmax$. We have numerically checked that the ground state of the Kitaev model remains in the flux-free sector provided that $C<\Cmax$.

\subsection{Landau levels and field inhomogeneities}

As a result of the field inhomogeneities, the clearest Landau-level signatures are obtained in the LDOS in the center of the system. This has been discussed in detail for graphene \cite{peeters13}, and our LDOS results including finite-size effects are similar to that in Ref.~\onlinecite{peeters13}.

\begin{figure}[t]
\centering
\includegraphics[scale=0.46]{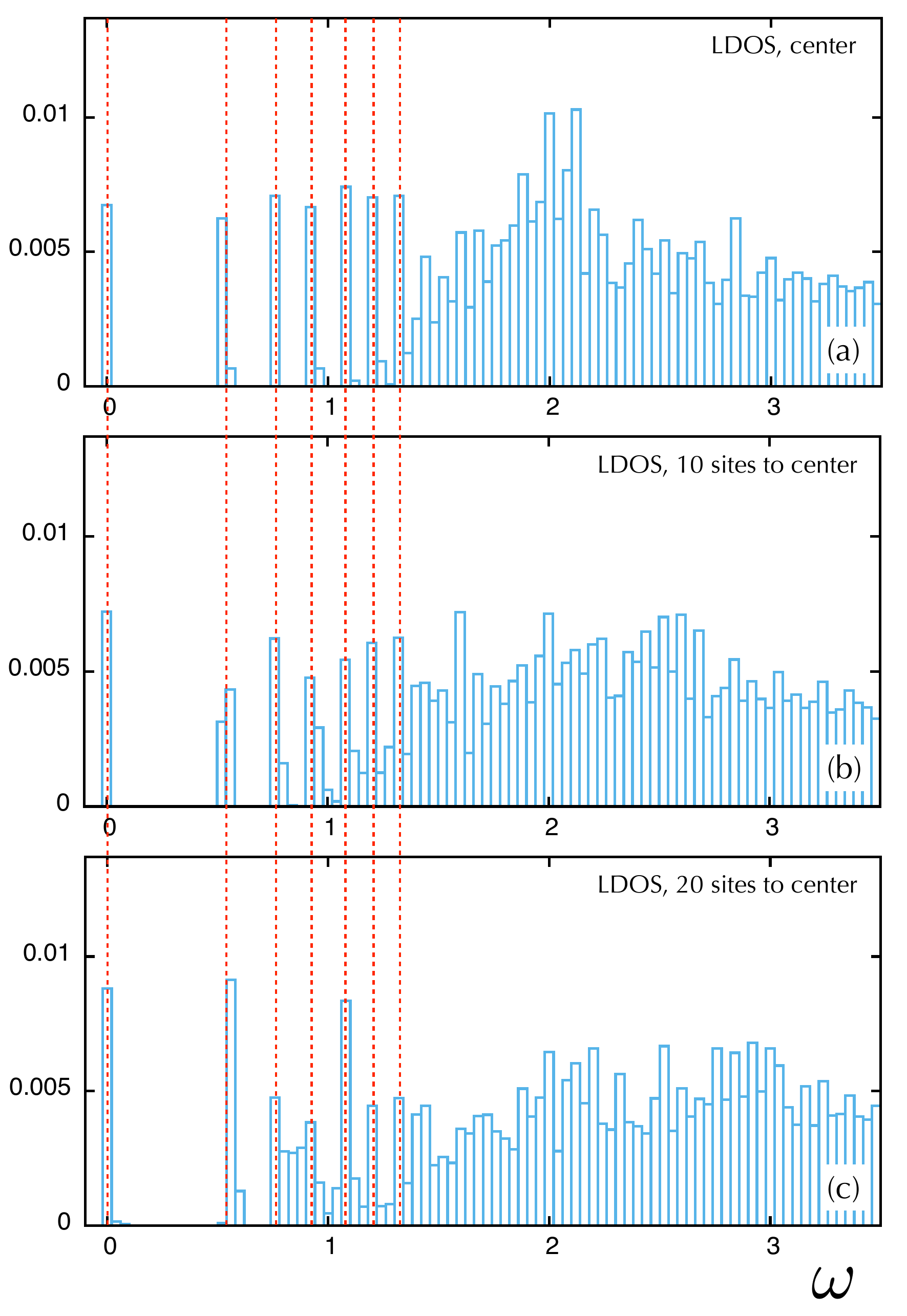}
\caption{LDOS in the center and off the center for $2N=15\,000$ spins and $C=0.004$.
(a) is identical to Fig.\,2\,(g), measured in the center. (b) and (c) show LDOS measured 10 and 20 sites away from center, respectively. Red dashed lines are guide-to-the-eye for comparing the position of the LLs.}
\label{fig:dos+ldos}
\end{figure}

We note that, despite the inhomogeneities, the pseudo-Landau levels display a large approximate degeneracy, i.e., the energy of the Landau-level peaks is essentially constant as function of spatial position in an extended region around the sample center. To illustrate this, we show in Fig.\,\ref{fig:dos+ldos} the LDOS of Fig.\,1\,g (main paper), measured at the sample center, and two corresponding LDOS further away from the center. In all three cases the Landau levels appear at the same energy. Close to the sample edge the energy position of the Landau levels slightly deviates (with the exception of the LLL). We note that larger values of $\beta$ reduce non-linearities in the strain pattern which are ultimately responsible for the deviations between Figs.\,\ref{fig:dos+ldos}\,(a), (b), and (c).

\subsection{Edge states}

While our system displays edge states at its zigzag edges, these are not topologically protected and do not energetically connect different Landau levels -- the latter can be clearly seen in our LDOS results for large values of strain $C$ (not shown). This is consistent with the fact that the strained system remains time-reversal invariant and hence is characterized by a vanishing Chern number.

\begin{figure}[t]
\centering
\includegraphics[scale=0.52]{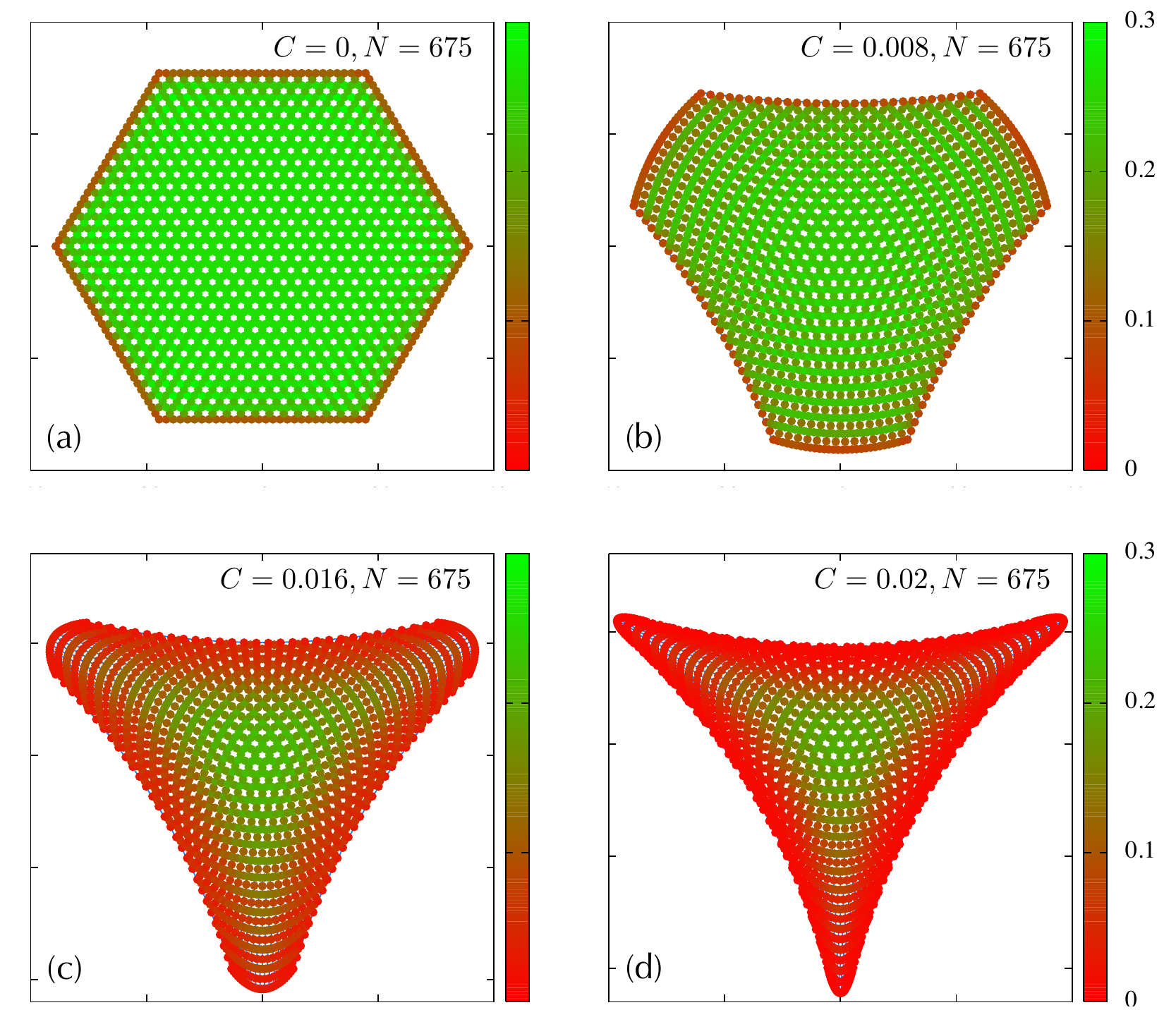}
\caption{Local flux gap $\Delta_{ij}/J$. The four panels correspond to different values of $C$:
(a) $C=0$,
(b) $C=0.008$,
(c) $C=0.016$, and
(d) $C=0.02$. The sample contains $2N=1350$ lattice sites.}
\label{fig:locgap}
\end{figure}

\subsection{Flux gap for inhomogeneous couplings}

For the homogeneous isotropic Kitaev model the flux gap, \ie the energy difference between the matter-fermion ground states in the two-flux and zero-flux sectors, is given by $\Delta=0.26J$ in the thermodynamic limit \cite{kitaev06}. Note that the lowest-energy two-flux states are those with the two fluxes being located in adjacent plaquettes.

Importantly, for a finite-size system with open boundaries the flux gap depends on the spatial position of the flux pair. This requires to introduce a {\em local} flux gap $\Delta_{ij}$ where $ij$ refers to the bond with $u_{ij}=-1$ adjacent to the fluxes. It is this local gap $\Delta_{ij}$ which represents the lower energy bound for the intermediate state $\ket{\lambda}$ in the spin correlator $S^{\alpha\alpha}_{ij}(\w)$.

In Fig.~\ref{fig:locgap} we show $\Delta_{ij}$ for different values of the strain $C$. Clearly, small and moderate values of $C$ lead to an essentially homogeneous gap in the interior of the system and a reduced gap near the edges only. In contrast, for large $C$ close to $\Cmax$ the gap profile becomes rather inhomogeneous, with a pronounced maximum in the middle. This can be easily understood: The strongly elongated bonds near the edges correspond to a reduced exchange coupling, and the energy cost of flipping a bond in such a plaquette is small.

\subsection{Local vs. zero-momentum spin correlations}

Given the inhomogeneous character of the system, with both the pseudo-magnetic field and the flux gap being approximately homogeneous in the interior only, one has to distinguish measurements in position and momentum space. Momentum-space spin correlations are probed by the dynamic structure factor $S(\vec{q},\w)$. For real-space measurements it is convenient to introduce a bond-local response function,
\begin{equation}
S_{\langle ij \rangle}(\w)=S^{\alpha\alpha}_{ii}(\w)+S^{\alpha\alpha}_{jj}(\w)+2S^{\alpha\alpha}_{ij}(\w)
\end{equation}
with $\alpha=\alpha_{ij}$. The structure factor is related to these local correlators by $\sum_\alpha S^{\alpha\alpha}(\vec{q}=0,\w) = 1/N \sum_{\langle ij\rangle} S_{\langle ij\rangle}(\w)$ where the sum $\langle ij\rangle$ is over all lattice bonds.

\begin{figure*}[t!]
\centering
\includegraphics[scale=0.66]{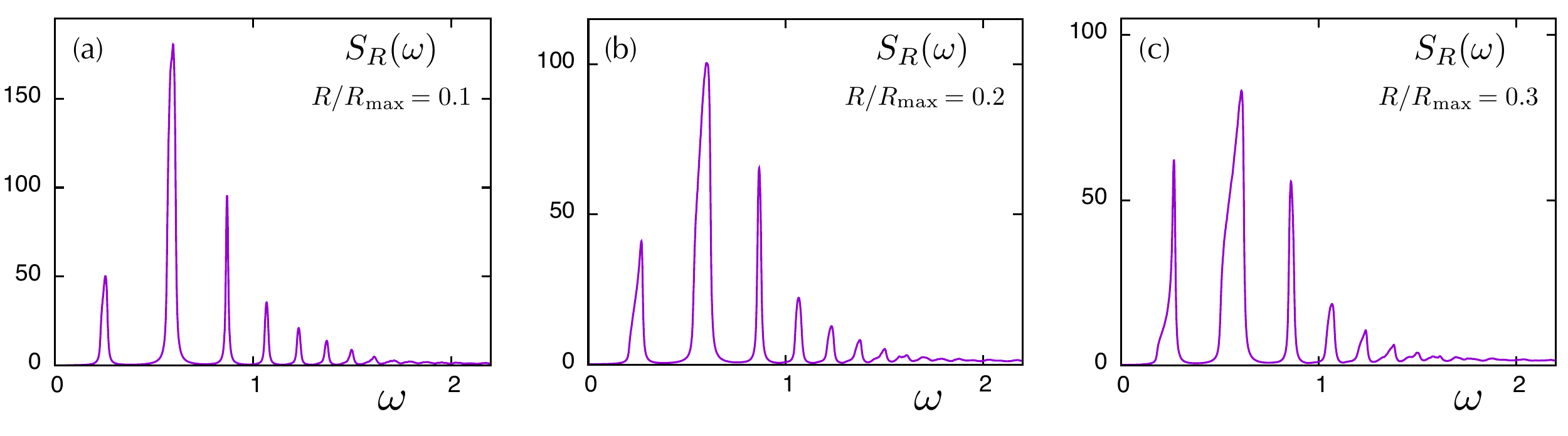}
\caption{
Correlators $S_R(\w)$ \eqref{srdef} -- representing a finite-area approximants to $S(\vec{q}=0,\w)$ -- for
(a) $R/R_{\rm max}=0.1$, featuring essentially the same peak structure as $S_{\langle ij \rangle}(\omega)$ in Fig.~\figmainres(i),
(b) $R/R_{\rm max}=0.2$, with visible broadening of the bound-state peak due to inhomogeneities, and
(c) $R/R_{\rm max}=0.4$.
The system parameters are $2N=15\,000$, $C=0.004$, and $R_{\rm max}$ is the radius of the smallest circle enclosing the entire flake (in unstrained coordinates). The Lorentzian broadening is $\gamma/J=0.005$.
}
\label{fig:fullSq}
\end{figure*}

The plots of Fig.~{\figmainres} in the main paper display local observables in the center of the system where the Landau-level signatures are clearest -- here isolated bound-state peaks are seen in $S_{\langle ij \rangle}(\w)$. Importantly, the condition of locality for the measurement can be relaxed: In Fig.~\ref{fig:fullSq} we show approximants to $(1/3)\sum_\alpha S^{\alpha\alpha}(\vec{q}=0,\w)$, defined as the sum of all correlators inside a circle of radius $R$ in the interior of the sample:
\begin{equation}
\label{srdef}
S_R(\w) = \frac{1}{N'} \sum_{\langle ij\rangle}^{r_i,r_j<R} S_{\langle ij \rangle}(\w)
\end{equation}
where $N'$ the number of bonds inside this circle. For small $R$, Fig.~\ref{fig:fullSq}(a), the result for $S_R$ resembles the single-bond result. Increasing $R$ leads to a smearing of the bound-state peaks -- this smearing primarily originates from the inhomogeneities in the flux gap.
Importantly, the bound-state peaks characteristic of the Landau-level structure are still clearly visible for $R$ values where the circle covers a sizeable fraction of the sample,
20\% in Fig.~\ref{fig:fullSq}(c). This shows that neutron-scattering measurements can detect the peak structure by measuring $S(\vec{q},\w)$ provided that the neutron beam is smaller than the sample size, such that the outer parts of the sample are not included in the probe.

\subsection{Thermodynamic limit}

The displacement pattern used to generate the pseudo-magnetic field, Eq.~{\eqdisplace}, cannot be extended to an infinitely large system, simply because bond lengths become negative. In other words, the thermodynamic limit of infinite linear size, $L/a_0\to\infty$, cannot be taken at fixed $\bar{C}$. We also note that, with increasing system size, the linearization in Eq.~{\eqinhomj} becomes invalid, which is eventually signalled by negative exchange couplings at the boundary which occur for $C>\Cmax$. Together, this raises the question whether there exists a well-defined limit where finite-size effects can be made small.

An interesting route is to take the combined limit $L/a_0\to\infty$ and $\bar{C}\to 0$, keeping $\bar{C}L/a_0$ (as well as the magnetoelastic factor $\beta$) fixed. In this limit, the pseudo-magnetic field scales to zero as $L/a_0\to\infty$, and with it the energetic spacing of the pseudo-Landau levels. As the ratio of system size and magnetic length is kept fixed, we speculate that the system approaches a scaling limit with well-defined Landau levels -- this will be investigated in future work.

\subsection{Spin correlations at finite temperature}

For an orbital magnetic field applied to graphene, the lowest Landau level (LLL) is located at exactly zero energy. For the present case of the strained Kitaev model, the LLL being at zero energy implies a highly degenerate many-body ground state. Notably, this degeneracy is lifted by finite-size effects -- all our mode energies $\e_m$ are positive.

For practical reasons, we evaluate the expectation value $\langle \ldots \rangle$ in the spin correlator, $S^{\alpha \beta}_{ij}$ in Eq.~{\eqsoft}, with the unique finite-system ground state $\ket{M_0}$, corresponding to strictly $T=0$. Importantly, our conclusions apply unchanged if $S^{\alpha \beta}_{ij}$ is calculated instead using a thermal average over the low-energy many-body states, i.e., at small finite temperature $T\ll \ELL_1$ where $\ELL_1$ is the energy of the first finite-energy Landau level. The argument proceeds as follows:
The relevant many-body initial states $\ket{M}$ involve any number of matter Majorana excitations from the LLL -- this precludes a full numerical evaluation of $S^{\alpha \beta}_{ij}$ at finite $T$. Importantly, for large systems, sizeable matrix elements $\bra{\lambda}\hat{c}_j\ket{M}$ are only obtained if the intermediate state contains, in addition to the excitation effectively created by $\hat{c}_j$, the {\em same} excitations as the initial state $\ket{M}$. This has been demonstrated in Ref.~\onlinecite{zschocke15} where the effect of the ground-state fermion-parity condition \cite{loss,zschocke15} on the spin structure factor has been examined: It was found that the two-particle continuum, obtained by starting from an initial state with one excited fermion, is virtually indistinguishable from the one-particle continuum, obtained by starting from an excitation-free initial state. More precisely, the differences between the two results for $S^{\alpha \beta}_{ij}$ scale with the inverse system size, the reason being that the wavefunctions of the additional excitations are spatially extended. Generalized to our system, this implies that all thermal contributions to the low-temperature spin correlator will have essentially the same peak structure, arising from the bound states between the Majorana Landau levels. While small finite temperatures will cause both small broadening and a small weight redistribution, the peak structure in $S(\w)$ shown in Fig.~{\figmainres} can be expected to be robust.

\begin{figure}[!t]
\centering
\includegraphics[scale=0.53]{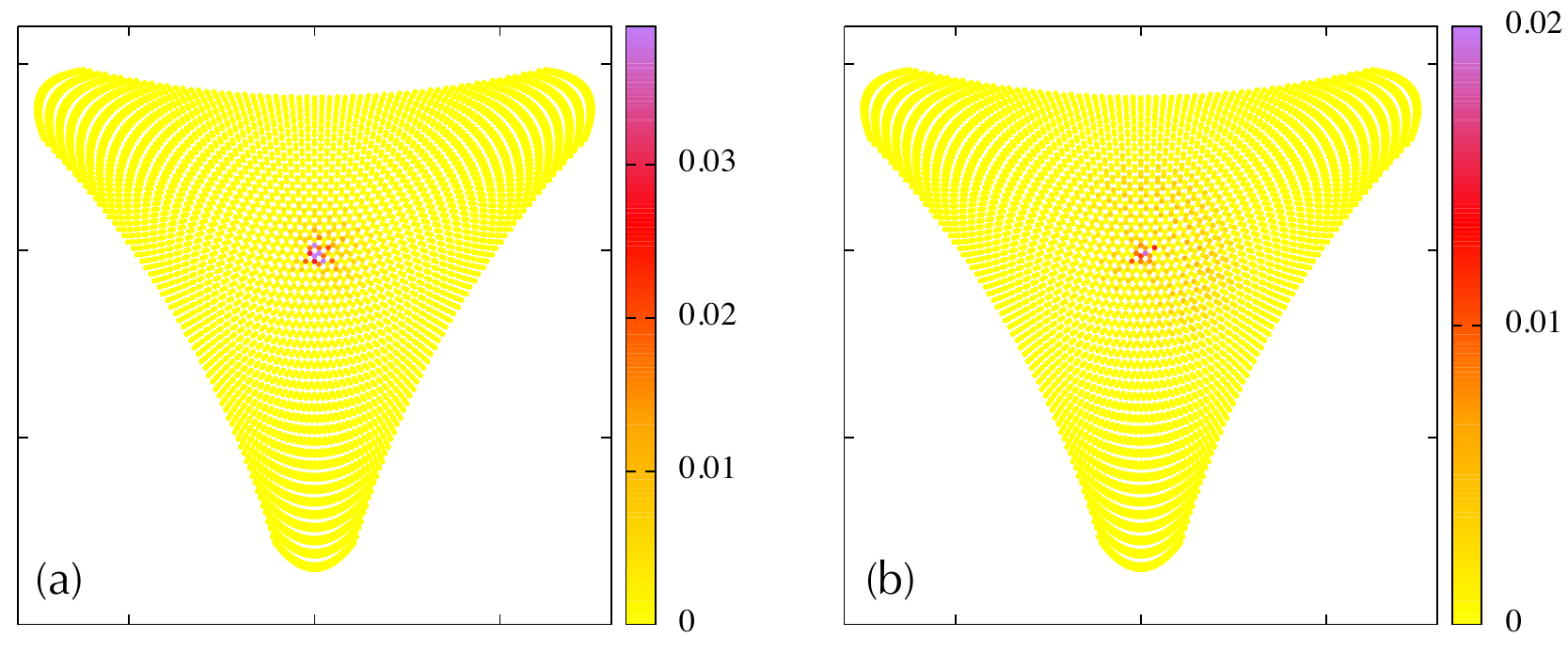}
\caption{Wavefunctions of the first two in-gap bound states in the two-flux sector for a system with $2N=5400$ spins and strain $C=0.0078$ where the flux pair has been created in the center.
(a) First in-gap state with energy $\EB_{n=1}\approx 0.44\,J$.
(b) Second in-gap state at $\EB_{n=2}\approx 0.86\,J$.
}
\label{fig:bswavefct}
\end{figure}


\section{Pseudo-Landau levels and bound states}

As emphasized in the paper, sufficiently large strain induces pseudo-Landau levels in the matter Majorana sector of the Kitaev model. At the same time, the dynamic spin correlator $S_{\langle ij \rangle}(\w)$ shows sharp $\delta$ peaks. Interestingly, these peaks (with the exception of the lowest one which can be attributed to the LLL in the two-flux sector) do {\em not} directly arise from the Landau levels. Instead the measurement dynamically generates an impurity leading to one localized bound state, causing a high-intensity peak in $S_{\langle ij \rangle}(\w)$, in each of the Landau-level gaps. Given that higher Landau levels tend to be smeared (because the analogy between strain and magnetic field is restricted to the low-energy limit), the spin correlator shows a much sharper peak structure than the matter-fermion DOS itself.

\subsection{Bound-state wavefunctions}

In Fig.~\ref{fig:bswavefct} we illustrate the nature of the probe-induced bound states in the two-flux sector, by plotting the probability distribution of eigenmode $m$ for A and B sites given by
\begin{equation}
w^m_{A,i} = U_{mi}^2/2\,,~~w^m_{B,j} = V_{mj}^2/2\,.
\end{equation}
The bound-state eigenmodes are highly localized near the flipped bond (which is taken to be in the center of the sample), which we have checked by computing the inverse participation ratio for these modes. The localization length increasing with increasing $n$, which is a simple result of the decreasing gap magnitude.

\begin{figure}[t]
\centering
\includegraphics[scale=0.7]{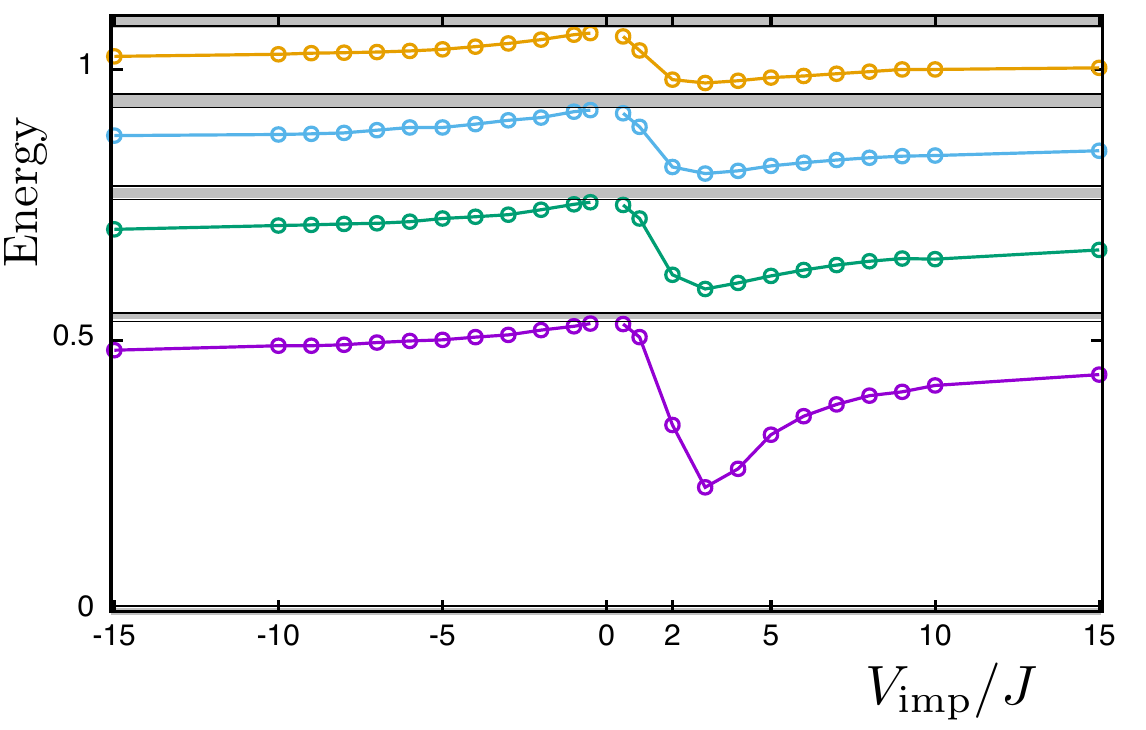}
\caption{Energy positions $\EB_n$ of the bound states caused by a bond impurity as a function of the impurity strength $V_{\rm imp}$. Note that $V_{\rm imp}=0$ is the clean case and $V_{\rm imp}=2J_{ij}$ corresponds to a flipped bond (``two-flux sector''). The data corresponds to $N=7500$ and $C=0.004$; the grey horizontal lines represent the energies of the pseudo-Landau levels.
}
\label{fig:genbound}
\end{figure}


\subsection{Bound states and topology}

A recent paper \cite{zaanen15} has proposed an interesting connection between topological states and local impurity physics, namely that the impurity-induced local in-gap Green's function can be used as diagnostic tool for topology. For time-reversal-invariant $\mathbb{Z}_2$ topological insulators in two space dimensions a local impurity always causes an in-gap bound state. This is because the real part of the local Green's function diverges at the lower edge of the upper band with opposite sign compared to the upper edge of the lower band. This guarantees the presence of an in-gap bound state. It was further shown that a topologically trivial insulator does not generally possess such in-gap bound states (but may do so in a finite interval of impurity strengths).

As our analysis has revealed the creation of localized in-gap states due to bond impurities, the latter representing a pair of Z$_2$ fluxes, we adapt the idea of Ref.~\onlinecite{zaanen15} to our setting. We consider a bond impurity of general strength, $\hat{V} = - \ii V_{\rm imp} \hat{c}_i \hat{c}_j$, in a honeycomb-lattice Majorana hopping problem subject to triaxial strain. While the impurity value $V_{\rm imp}=-2$ (measured in units of $J_{ij}$) corresponds to a flipped bond, \ie a two-flux sector, here we consider arbitrary values of $V_{\rm imp}\in[-15,15]$, with the clean case corresponding to $V_{\rm imp}=0$.
As shown in Fig.~\ref{fig:genbound}, we find that bond impurities of any strength induce bound states between the Landau levels (except for the immediate vicinity of $V_{\rm imp}=0$ where any bound state must occur close to a gap edge and is thus hard to detect in a finite-size system). Hence, the criterion of Ref.~\onlinecite{zaanen15} suggests that the matter-fermion sector of the strained Kitaev model displays non-trivial topology.
